\documentclass{svproc}
\usepackage{graphicx}

\usepackage{url}

\begin{document}
\mainmatter              % start of a contribution
\title{Hot-Get-Richer Network Growth Model}
\titlerunning{Hot-Get-Richer Network Growth Model}  % abbreviated title (for running head)
\author{Faisal Nsour\inst{1} \and Hiroki Sayama\inst{1,2}}
\authorrunning{Faisal Nsour et al.} % abbreviated author list (for running head)
\tocauthor{Faisal Nsour, Hiroki Sayama}
\institute{Department of Systems Science and Industrial Engineering\\
Binghamton University, Binghamton, NY, USA \\
\email{fnsour1@binghamton.edu} \\
\email{sayama@binghamton.edu} \\
\and
Waseda Innovation Lab, Waseda University, Japan} 

\maketitle              % typeset the title of the contribution

\begin{abstract}
Under preferential attachment (PA) network growth models late arrivals are at a disadvantage with regard to their final degrees. Previous extensions of PA have addressed this deficiency by either adding the notion of node fitness to PA, usually drawn from some fitness score distributions, or by using fitness alone to control attachment. Here we introduce a new dynamical approach to address late arrivals by adding a recent-degree-change bias to PA so that nodes with higher relative degree change in temporal proximity to an arriving node get an attachment probability boost. In other words, if PA describes a rich-get-richer mechanism, and fitness-based approaches describe good-get-richer mechanisms, then our model can be characterized as a hot-get-richer mechanism, where hotness is determined by the rate of degree change over some recent past. The proposed model produces much later high-ranking nodes than the PA model and, under certain parameters, produces networks with structure similar to PA networks.
\keywords{Preferential attachment, network growth, first-mover advantage, degree dynamics, winner-take-all, hot-get-richer.}
\end{abstract}
\section{Introduction}

Under the preferential attachment (PA) growth model, late-arriving nodes are at a disadvantage with regard to their final degree. To account for high-degree later arrivals that are often observed in real-world empirical networks, several methods of extending or replacing the PA growth model have been explored, often with some sort of node fitness either replacing or modifying PA [1]. Some approaches use node age as a key factor to determine node fitness [4][5][7]. Node extinction is an extreme form of fitness-based growth that accounts for nodes becoming ineligible for new attachment by aging out of the pool of potential attachments for arriving nodes [10]. Fitness-based growth has been found to produce so-called scale-free networks even in the absence of preferential attachment [2]. It is interesting to note that so long as ranking information is preserved, fitness values themselves need not be available to arriving nodes [3]. 

Comparatively fewer studies have focused on the late arrivals themselves [4][9]. As opposed to tying growth to fitness measures, the study by Mokryn et al. [9] proposes a new model called Trendy Preferential Attachment (TPA) where attachment probability is a function of decaying edge relevance, allowing high-degree nodes to wane in their ability to attract new edges thereby opening the way for late arrivals to grow in importance.

In this paper we propose a new degree-recency-biased preferential attachment (DRPA) growth model. The DRPA model balances relative recent degree change against overall degree to determine the attachment probability. Furthermore, we explore not only the final network structure resulting from the growth model, but also the qualitative changes in rank of high-ranking late arrivals in particular as the network grows. DRPA models the creation of high-ranking nodes from late-arriving nodes by adding the concept of degree recency to classical PA. The intuition is simple: late-arriving nodes can become attractive because they have, in some recent past, been connected to at a higher rate relative to their degree. Recent degree change signals potential future growth which biases PA growth. To analyze results, and compare DRPA to PA in terms of late arrivals, we also introduce measures of arrival position and degree change trajectory in growing networks.

The DRPA model is similar to the TPA model in that it accounts for shifting regimes of degree distribution (i.e., not only degree distribution but which nodes gain or lose rank) based on temporal dynamics. However, the two models approach trend determination in different ways. TPA determines the attachment probability of a new node to an existing node $i$ as a time-weighted sum  $p(i) =  f(1)k_i(t - 1) + f(2)k_i(t - 2) + ... + f(t - 1)k_i(1)$, where $f(t)$ is a monotonically decreasing function. The function $f(t)$ in TPA can change between networks (and even be monotonically increasing to provide conservation of degree history), and TPA requires the node's full history to be available at each attachment. In contrast, DRPA (see eq. 1 below) relies only on the degree of node $i$ at arrival time and the most degree change. These differences suggest that while similar in intent, the models will likely produce differing results, and in some cases (e.g., due to the history requirement of TPA) may not be interchangeable. We note the similarities and differences and recognize that a follow up in-depth comparison is needed to better understand effect of temporal dynamics on change in degree distribution.

In real-world networks, fitness is indeed often intrinsic to each node. A new website, for example, may offer a unique value to users and thus gain inbound links at a much higher rate than its age-proportional rate would dictate. However, even with inherent fitness, recent changes in topology may exert an influence. Some product may be interesting to consumers simply because it has suddenly been purchased disproportionately to its previous sales than have competing products, even long established ones. In a word, it has become hot. If preferential attachment is a rich-get-richer mechanism, and fitness-based approaches are good-get-richer mechanisms, then we might say that DRPA (and related models) is a hot-get-richer mechanism wherein late-movers may have an advantage over first movers due to their increased likelihood of higher relative degree change.
\section{Degree-Recency-Biased Preferential Attachment Model}

The DRPA model can be thought of as being PA in which node degrees are weighted by their relative recent changes used as a multiplicative factor. The model is denoted as follows, with an arriving node having probability of attaching to existing node \emph{i}:

\begin{equation}
  p_i(t) = \frac{k_i(t)R_i(t)}{\sum_{j}^{t} k_j(t)R_j(t)}
\end{equation}
Here $k_i(t)$ is the degree of node \emph{i}, and $R_i(t)$ is the recency attachment factor of node \emph{i}, defined as

\begin{equation}
R_i(t) =  \left(\frac{k_i(t) - k_i(t - r)}{k_i(t)}\right)^\beta ,
\end{equation}
where \emph{r} is the recency span parameter indicating how far back in the network’s growth cycle to determine a given node’s degree change. Parameter $\beta$ is a tuning parameter, on the unit interval, that controls how much the model should prefer absolute degree $k_i(t)$ or recent change in degree $(k_i(t) - k_i(t-r))$. When $\beta$ goes to 0, $p_i(t)$ becomes classical preferential attachment. As it goes to 1, $p_i(t)$ goes to an attachment based only on recent degree change, defined as:

\begin{equation}
p_i(t) =  \frac{k_i(t) - k_i(t - r)}{\sum_{j}^{t}k_j(t) -  k_j(t - r)}
\end{equation}

A warm up period is allowed so that at any given moment Eq.\ (3) is valid. Negative values of $\beta$ would reward the opposite of hotness, what we might call ``conservativeness'' (i.e., slow growth of high-degree nodes), but these ranges were not explored in the present study. %To summarize the above, the model (1) scales preferential attachment by recent degree change relative to degree.

We next attempt to obtain an analytical understanding of the DRPA model dynamics. For this purpose, we assume that the time is continuous (even though the numerical simulations described later were conducted with discrete time) and the recent degree change $(k_i(t)-k_i(t-r))$ can be approximated by a separate dynamical variable called recency, denoted as $r_i(t)$, that decays exponentially with time but grows by $k_i'(t)$. The latter assumption allows us to avoid using time-delay differential equations and thus significantly simplifies the analytical work.

With these continuous-time assumptions, we describe the dynamics of the model as

\begin{equation}
	k_i'(t) = m p_i(t) = m \frac{k_i(t)^{1-\beta} r_i(t)^{\beta}}{\displaystyle\sum_{j<t}k_j(t)^{1-\beta}r_j(t)^{\beta}},				       
\end{equation}
\begin{equation}
	r_i'(t) = -\alpha r_i(t) + k_i'(t),
\end{equation}
for $i < t$, where $m=\sum_{i<t}k_i'(t)$ is the rate of total degree growth for nodes $i<t$ (this corresponds to the number of edges by which a newcomer node $i=t$ is connected to the network), $p_i(t)$ is the preferential selection probability of node \emph{i}, and $\alpha$ is the parameter that determines the exponential decay rate of recency. 

We then consider an asymptotic scenario in which the preferential selection probability distribution converges to a well-defined time-invariant distribution $\bar{p}_i$ as $t\to\infty$, i.e.,

\begin{equation}
	k_i'(t) = m \bar{p}_i.		           
\end{equation}

Solving Eqs.\ (5) and (6) gives

\begin{equation}
	k_i(t) =m \bar{p}_i t+C_{1,i},
\end{equation}
\begin{equation}
    r_i(t)=\frac{m \bar{p}_i}{\alpha}+C_{2,i} \, e^{-\alpha t}.
\end{equation}

Now we plug these back into the definition of $p_i(t)$ (Eq.\ (4)) to obtain the shape of $p_i$:

\begin{eqnarray}
  \bar{p}_i	&=& \frac{
    \left(m \bar{p}_i t+C_{1,i}\right)^{1-\beta}
    \left(\displaystyle\frac{m \bar{p}_i}{\alpha} + C_{2,i}\,e^{-\alpha t}\right)^\beta
    }
       {\displaystyle\sum_{j<t}{\left(m \bar{p}_i t+C_{1,j}\right)^{1-\beta}
       \left(\displaystyle\frac{m \bar{p_j}}{\alpha} + C_{2,j}\,e^{-\alpha t}\right)^\beta}} \\
&=&\frac{\left(m \bar{p}_i t\right)^{1-\beta}
  \left(1+\displaystyle\frac{C_{1,i}}{m \bar{p}_i t}\right)^{1-\beta}
  \left(\displaystyle\frac{m \bar{p}_i}{\alpha}\right)^\beta
  \left(1+\displaystyle\frac{\alpha C_{2,i}\,e^{-\alpha t}}{m \bar{p}_i}\right)^\beta
  }{\displaystyle\sum_{j<t}
  {\left(m \bar{p_j} t\right)^{1-\beta}
  \left(1+\displaystyle\frac{C_{1,j}}{m \bar{p_j} t}\right)^{1-\beta}
  \left(\displaystyle\frac{m \bar{p_j}}{\alpha}\right)^\beta
  \left(1+\displaystyle\frac{\alpha C_{2,j}\,e^{-\alpha t}}{m \bar{p_j}}\right)^\beta}}  \\
&=& \frac{\bar{p}_i
     \left(1+\displaystyle\frac{C_{1,i}}{m \bar{p}_i t}\right)^{1-\beta}
     \left(1+\displaystyle\frac{\alpha C_{2,i}\,e^{-\alpha t}}{m \bar{p}_i}\right)^\beta }
  {\displaystyle\sum_{j<t}
  {\bar{p_j}
  \left(1+\displaystyle\frac{C_{1,j}}{m \bar{p_j} t}\right)^{1-\beta}
  \left(1+\displaystyle\frac{\alpha C_{2,j}\,e^{-\alpha t}}{m \bar{p_j}}\right)^\beta}}  \\
  &\approx&
  \frac{\bar{p}_i
     \left(1+\left(1-\beta\right)\displaystyle\frac{C_{1,i}}{m \bar{p}_i t}\right)
     \left(1+\beta\displaystyle\frac{\alpha C_{2,i}\,e^{-\alpha t}}{m \bar{p}_i}\right) }
  {\displaystyle\sum_{j<t}
  {\bar{p_j}
  \left(1+\left(1-\beta\right)\displaystyle\frac{C_{1,j}}{m \bar{p_j} t}\right)
  \left(1+\beta\displaystyle\frac{\alpha C_{2,j}\,e^{-\alpha t}}{m \bar{p_j}}\right)}}
\end{eqnarray}

By expanding the expressions and ignoring higher-order small terms (fractional parts inside the parentheses), we obtain

\begin{eqnarray}
\bar{p}_i &\approx& 
 \frac{\bar{p}_i + \left(1-\beta\right)\displaystyle\frac{C_{1,i}}{m t} + \beta\displaystyle\frac{\alpha C_{2,i}\,e^{-\alpha t}}{m}}
 {\displaystyle\sum_{j<t}{\left(\bar{p_j} + \left(1-\beta\right)\displaystyle\frac{C_{1,j}}{m t} + \beta\displaystyle\frac{\alpha C_{2,j}\,e^{-\alpha t}}{m}\right)}}\\
 &=&
 \frac{\bar{p}_i + \displaystyle\frac{1-\beta}{m t} C_{1,i} + \displaystyle\frac{\alpha \beta e^{-\alpha t}}{m}C_{2,i}}
 {1 + \displaystyle\frac{1-\beta}{m t}C_1 
 + \displaystyle\frac{\alpha\beta e^{-\alpha t}}{m} C_2},
\end{eqnarray}
where $C_1=\sum_i{C_{1,i}}$ and $C_2=\sum_i{C_{2,i}}$. Solving Eq.\ (14) for $\bar{p}_i$ gives

\begin{equation}\label{eq:final}
\bar{p}_i = \frac{C_{1,i}\,e^{\alpha t} (1-\beta) + C_{2,i}\,\alpha \beta t}
{C_{1}\,e^{\alpha t} (1-\beta) + C_{2}\,\alpha \beta t}.
\end{equation}

Eq.\ (\ref{eq:final}) presents several analytical predictions:

\begin{itemize}
  \item If $\beta\to 0$,  $\bar{p}_i\approx C_{1,i}/C_1$, i.e., the degree-based preference dominates. This makes the model identical to traditional degree-based preferential attachment.
  \item If $\beta\to 1$,  $\bar{p}_i\approx C_{2,i}/C_2$, i.e., the recency-based preference dominates.
  \item For $0<\beta<1$, eventually $\bar{p}_i\approx C_{1,i}/C_1$, i.e., the degree-based preference dominates in the long run.
  \item $ |C_2 \alpha \beta t| > |C_1 e^{\alpha t}(1 - \beta)| \;\; \exists t > 0$ if and only if $\beta > \frac{C_1 e}{C_1 e + C_2}$. This means that the recency-based dynamics can play a significant role for a finite period of time only if $\beta$ is above a certain threshold.
\end{itemize}

\section{Simulation}

\subsection{Experimental Settings}

Each simulation run grew a new model network to 1,000 nodes. The degree recency bias (parameter $\beta$) took values on the set [0.0, 0.1, 0.2, 0.25, 0.3, 0.4, 0.5, 0.6, 0.7, 0.8, 0.9, 1.0], and the degree recency range (parameter \emph{r}) took values on the set [5, 20, 50, 100]. The product of those sets define all possible model configurations. Each configuration was used for 50 independent simulation runs.

\subsection{Measurements}

The basic measurements used to compare models are based on node \emph{arrival} and node degree \emph{rank}. Arrival is the node’s ordinal position in the overall node arrival sequence (e.g., the 10th arriving node has arrival number of 10). Degree rank is a node’s position in a network’s degree centrality (e.g., the highest degree node has rank of 1). Based on these basic concepts, we propose two additional composite measures specific to late-arrival analysis:

\subsubsection{Rank-Arrival Difference.} The difference between a node’s final rank and its arrival number gives the relative magnitude of change of rank. For example, if the 100th arriving node is 10th in order of final degree, its rank-arrival difference is 90. 

\subsubsection{Rank Change Index.} We introduce the Rank Change Index (RCI) measure to quantity the amount of rank change in a network.  

\begin{equation}\label{eq:rci}
RCI = \frac{\displaystyle\sum_n{\mid n - Rank_n \mid}}
           {\displaystyle\sum_{m=1}^{m=N}{ \mid m - (N - m) \mid}}  
\end{equation}
where \emph{N} is the total node count, and \emph{n} is a given node’s arrival number. RCI sums rank-arrival absolute differences for all nodes and normalizes that by the  rank-arrival absolute differences of an idealized, perfectly reversed network.
    A perfect reversal of order (i.e., where the last arrival is ranked first, next to last is second, and so on) results in an RCI of 1. At the other extreme, a perfect alignment of arrival and rank (i.e., first arriving node has rank one and so on) results in an RCI of 0.

\subsection{Results}

Simulation results are described in the following sections on overall rank change, rank change of highest degree nodes, network degree distribution, and parameter influence.

To provide some intuition for the results, we first provide an position-averaged list of the top 1 percent highest degree nodes:

\begin{table}[tb]
\centering
 \caption{List of top 10 nodes (by degree) with their arrival number averaged. The DRPA row contains arrivals averaged over all $\beta$ values except 0. }
 \begin{tabular}{l l} 
 \hline
 Model & {Top 10 (ordered by degree) Arrivals Averaged }\\ [0.5ex] 
 \hline\hline
 DRPA &  [122.23, 186.85, 217.8, 235.4, 244.64, 253.15, 268.06, 278.55, 280.25, 284.01]\\
  ($\beta>0$) & \\
 \hline
 PA & [2.15, 5.46, 10.41, 14.63, 16.66, 21.62, 24.3, 26.68, 28.57, 32.79] \\
 ($\beta=0$) &\\
 \hline
\end{tabular}
\label{top_ten}
\end{table}

The data in table \ref{top_ten} show a clear difference between the two models in terms of how late late-arriving nodes can be in top degree positions. The mean arrival in the DRPA row is 237.09, and is 18.33 in the PA row. 

\begin{figure}[t]
\centering
%\begin{subfigure}{.6\textwidth}
%  \centering
  \includegraphics[width=.73\textwidth]{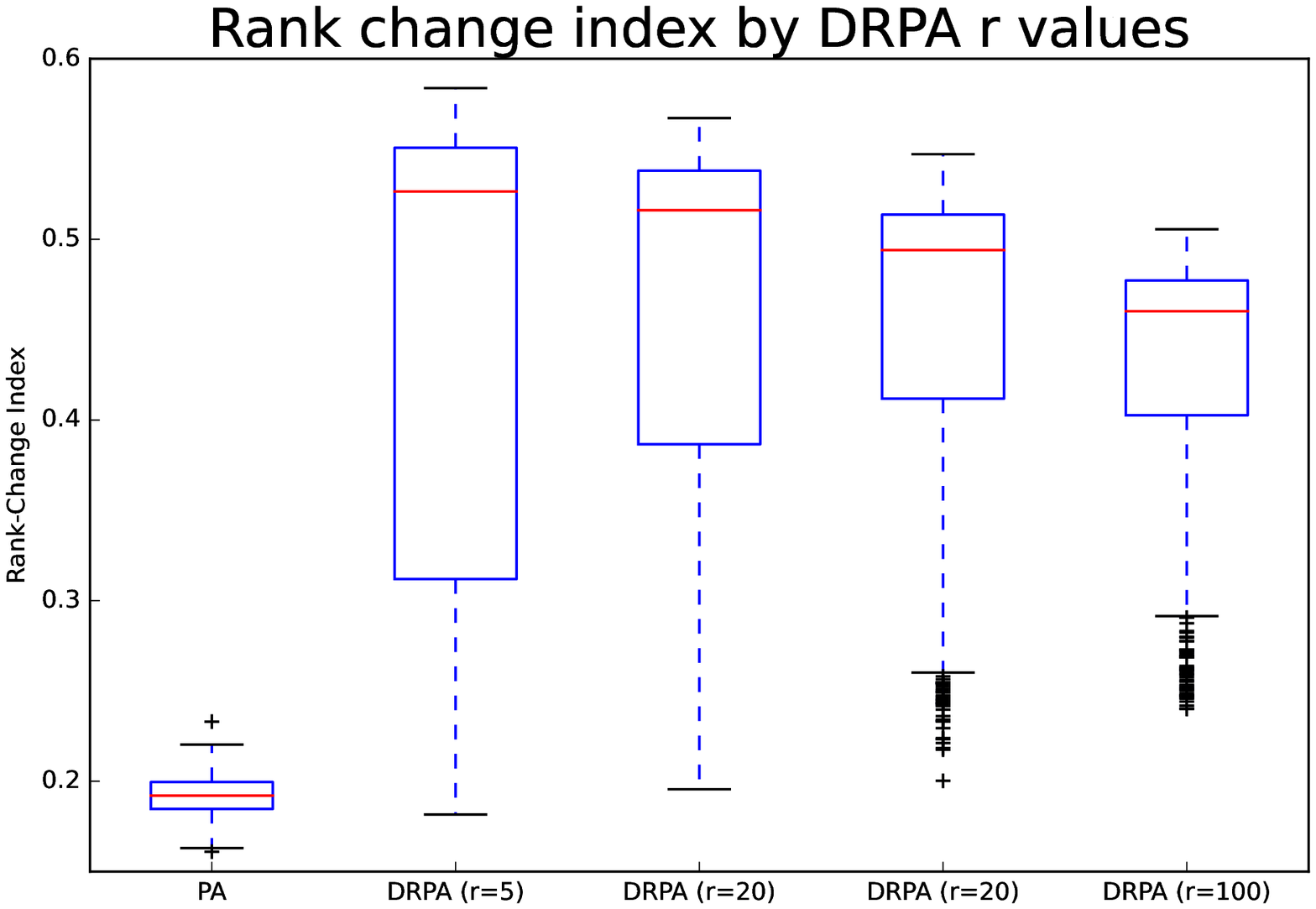}
%  \label{fig:rci_sub1}
%\end{subfigure}%
%\begin{subfigure}{.6\textwidth}
%  \centering
  \includegraphics[width=.73\textwidth]{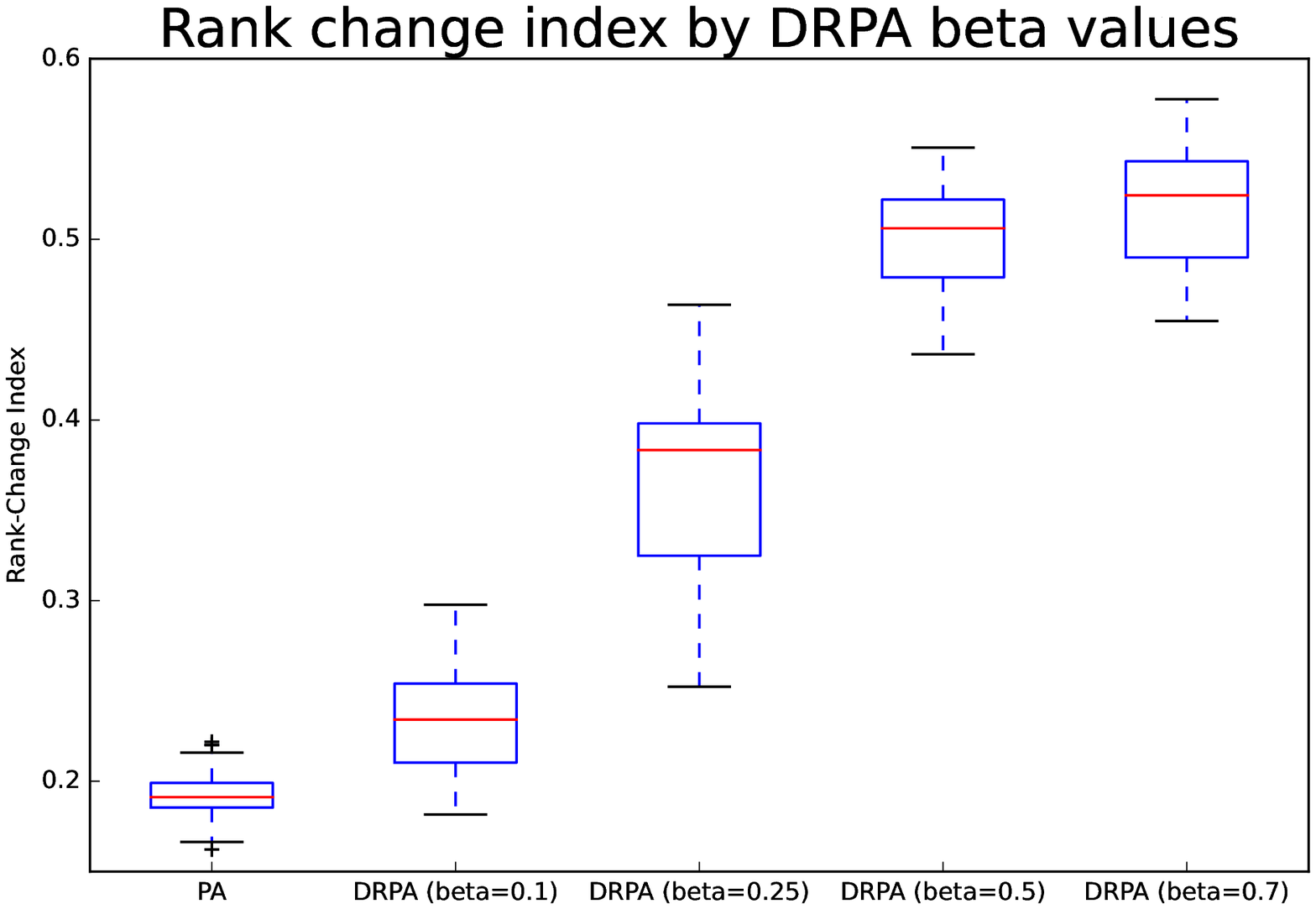}
%  \label{fig:rci_sub2}
%\end{subfigure}
%\setcounter{figure}{0}
\caption{Rank-change index by $r$ values (top), and $\beta$ values (bottom). The top pane contains all $\beta$ values, segmented by \emph{r}. The bottom pane contains all \emph{r} values, segmented by $\beta$ in set [0.1, 0.25, 0.5, 0.7]}

\label{fig:rci}
\end{figure}

\begin{figure}[t]
\centering
%\begin{subfigure}{.6\textwidth}
%  \centering
  \includegraphics[width=.73\textwidth]{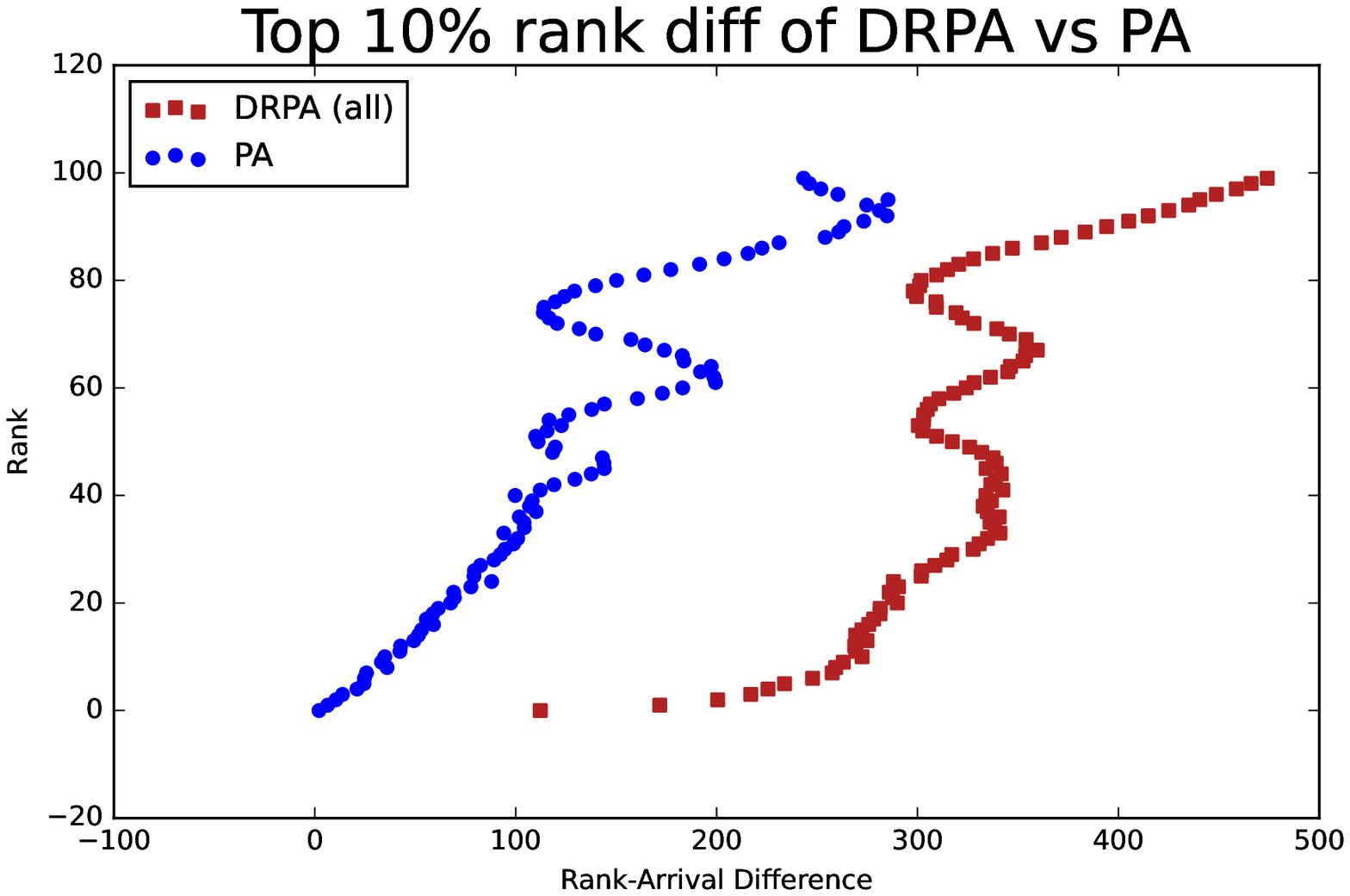}
%  \label{fig:rank_sub1}
%\end{subfigure}%
%\begin{subfigure}{.6\textwidth}
%  \centering
  \includegraphics[width=.73\textwidth]{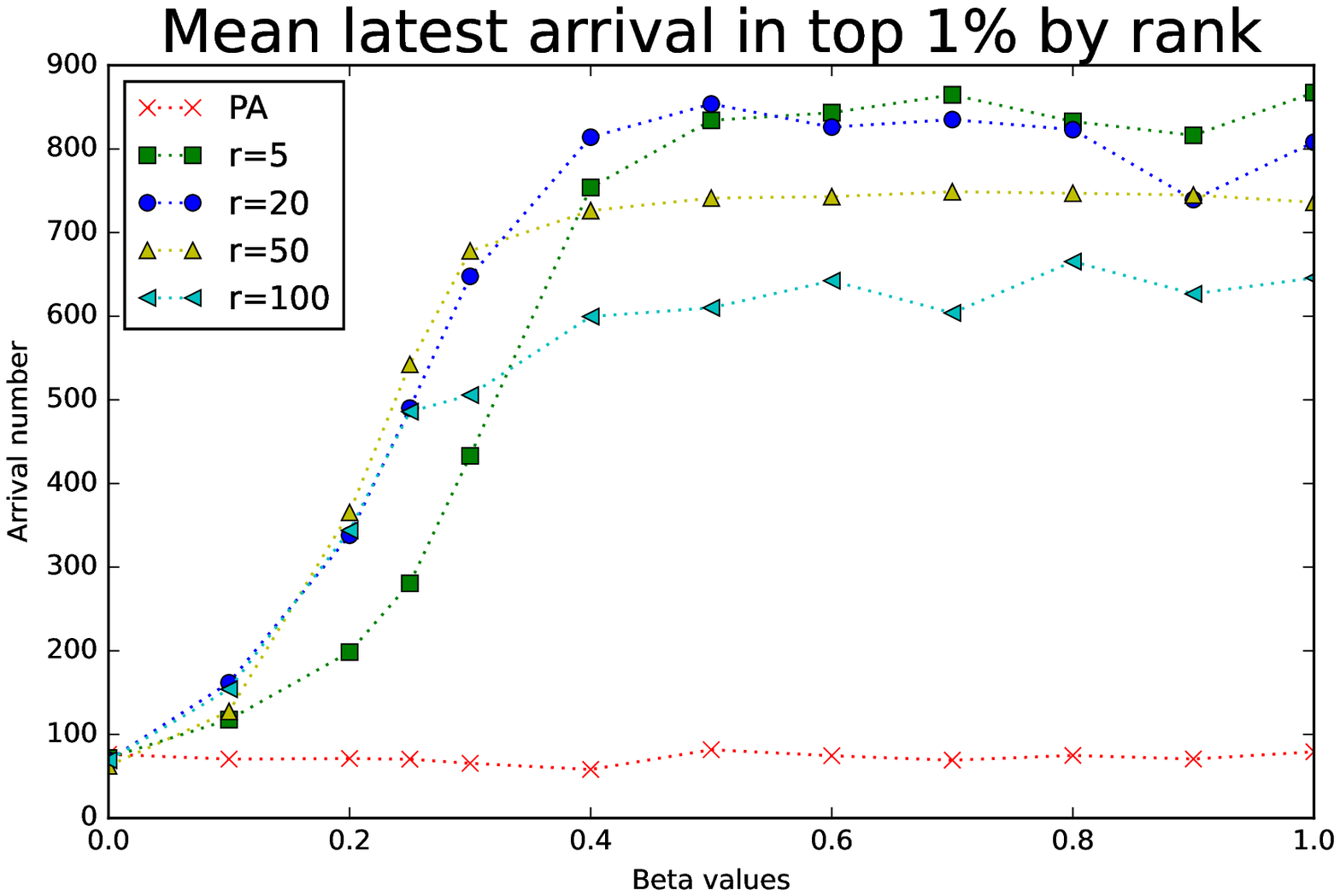}
%  \label{fig:rank_sub2}
%\end{subfigure}
%\setcounter{figure}{1}
\caption{Rank-arrival difference for top 10 percent, by rank (top). Max arrival in top 1 percent, by rank (bottom).}

\label{fig:rank_arrival}
\end{figure}

\subsubsection{Overall Rank Change.} Applying RCI in Eq.\ (\ref{eq:rci}) to simulation results produces the results in Fig. \ref{fig:rci} above. Results were separated by \emph{r} values (top) and $\beta$ values (bottom). The PA model shows a narrow range of RCI measurements, centering just below 0.2, and indicating that there is some amount of rank change occurring even in PA networks because of stochasticity in their growth processes. In contrast, the DRPA networks show markedly higher RCI, with a clear upward trend as $\beta$ increases (fig. \ref{fig:rci}, bottom). 

\subsubsection{Top Rank Comparison.} 
\begin{figure}[tb]
\centering
 \includegraphics[width=.8\textwidth]{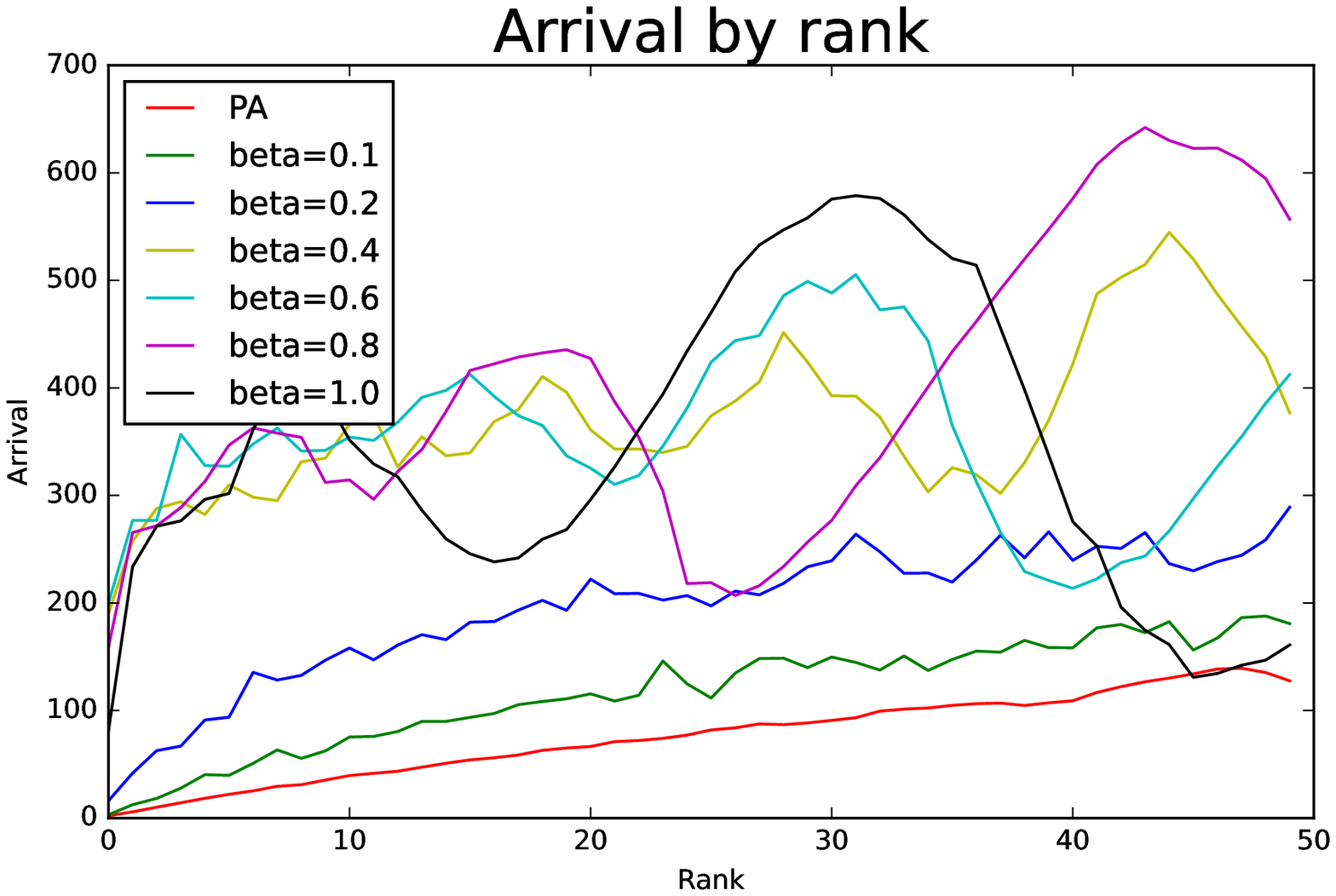}
 \caption{Node arrival sequence number by node final degree rank (top 5 percent, by degree rank)}
 \label{fig:arrival_by_rank}
\end{figure}

In Fig.\ \ref{fig:rank_arrival} (top) we show rank-arrival difference scores for the top 10 percent of nodes, by degree. The x-axis is rank-arrival difference, and the more to the right scores are indicates later arrivals. Fig.\ \ref{fig:rank_arrival} (bottom) shows the maximally latest arrival by the various \emph{r} parameter values. At lowest $\beta$ value of 0, the latest arrivals in DRPA and PA align. As $\beta$ increases, so does the latest arriving node's arrival number. Fig. \ref{fig:arrival_by_rank} shows change of degree rank by arrival number, by the various $\beta$ values. Note that above $\beta=0.2$, arrival numbers grow very quickly indicating that much later arrivals are finding their way into top positions (by degree). For the higher $\beta$ models, the very late arrivals are actually a sign of the flattening of the degree distribution, detailed in the following section.

\subsubsection{Network Degree Structure.}
We find that not only is lateness influenced by $\beta$, but so is the structure of the resulting networks. In particular, we observe the following:

\begin{table}[tb]
\centering
 \caption{Generated network degree statistics. All have mean degree near 2.0, and minimum degree of 1.}
 \begin{tabular}{c c c c} 
 \hline
 Network & Deg. Max. & Deg. Variance & Deg. Skewness \\ [0.5ex] 
 \hline\hline
 PA & 37.07 & 5.34 & 9.07 \\ 
 \hline
 DRPA($\beta$ = 0.0) &  37.60 & 5.27 & 8.89 \\
 \hline
 DRPA($\beta$ = 0.1) &  52.55 & 8.22 & 12.90 \\
  \hline
 DRPA($\beta$ = 0.2) &  47.82 & 4.74 & 14.06 \\
  \hline
 DRPA($\beta$ = 0.3) &  13.19 & 0.49 & 8.19 \\
  \hline
 DRPA($\beta$ = 0.4) &  7.76 & 0.15 & 7.25 \\
  \hline
 DRPA($\beta$ = 0.5) &  7.03 & 0.10 & 7.42 \\
  \hline
 DRPA($\beta$ = 0.6) &  6.37 & 0.08 & 7.45 \\
  \hline
 DRPA($\beta$ = 0.7) &  6.62 & 0.08 & 7.86 \\
  \hline
 DRPA($\beta$ = 0.8) &  5.93 & 0.07 & 6.74 \\
  \hline
 DRPA($\beta$ = 0.9) &  5.83 & 0.06 & 7.22 \\
  \hline
 DRPA($\beta$ = 1.0) &  5.80 & 0.06 & 7.17 \\
 \hline
\end{tabular}
\label{var_table}
\end{table}

Table \ref{var_table} shows a trend towards a relatively flat degree distributions as $\beta$ grows above the critical 0.2 to 0.25 region. Below that value, degree variance and higher skewness indicate the presence of the super nodes, that is, a few (possible just one) nodes with extremely high degree. At $\beta=0$, we observe a return of degree distribution to the PA model. This apparent transition of behaviors after $\beta = 0.2$ agrees with the analytical prediction derived in the previous section. 

\begin{figure}[tb]
\centering
%\begin{subfigure}{.5\textwidth}
%  \centering
  \includegraphics[width=.496\textwidth]{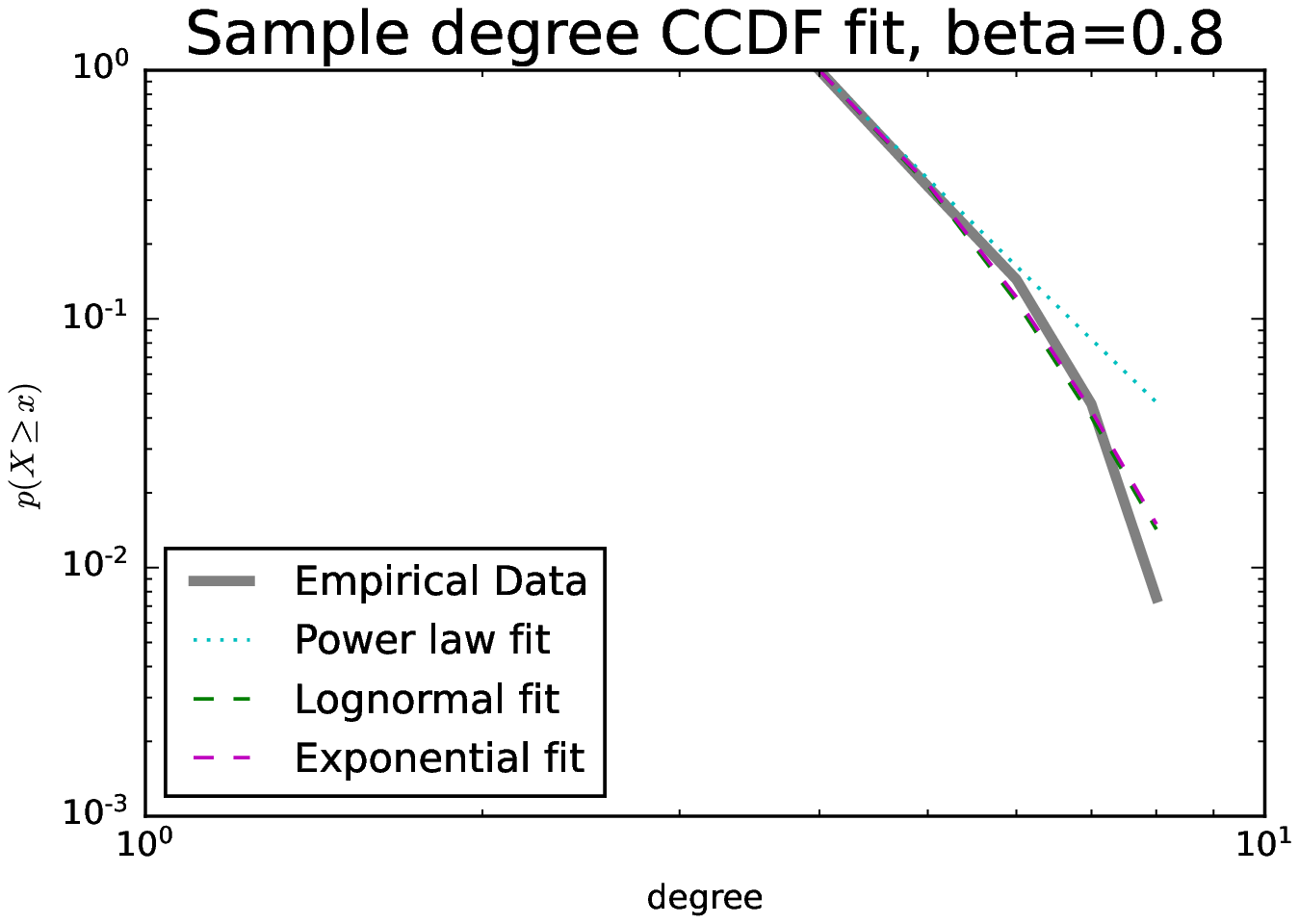}
%  \label{fig:rci_sub1}
%\end{subfigure}%
%\begin{subfigure}{.5\textwidth}
%  \centering
  \includegraphics[width=.496\textwidth]{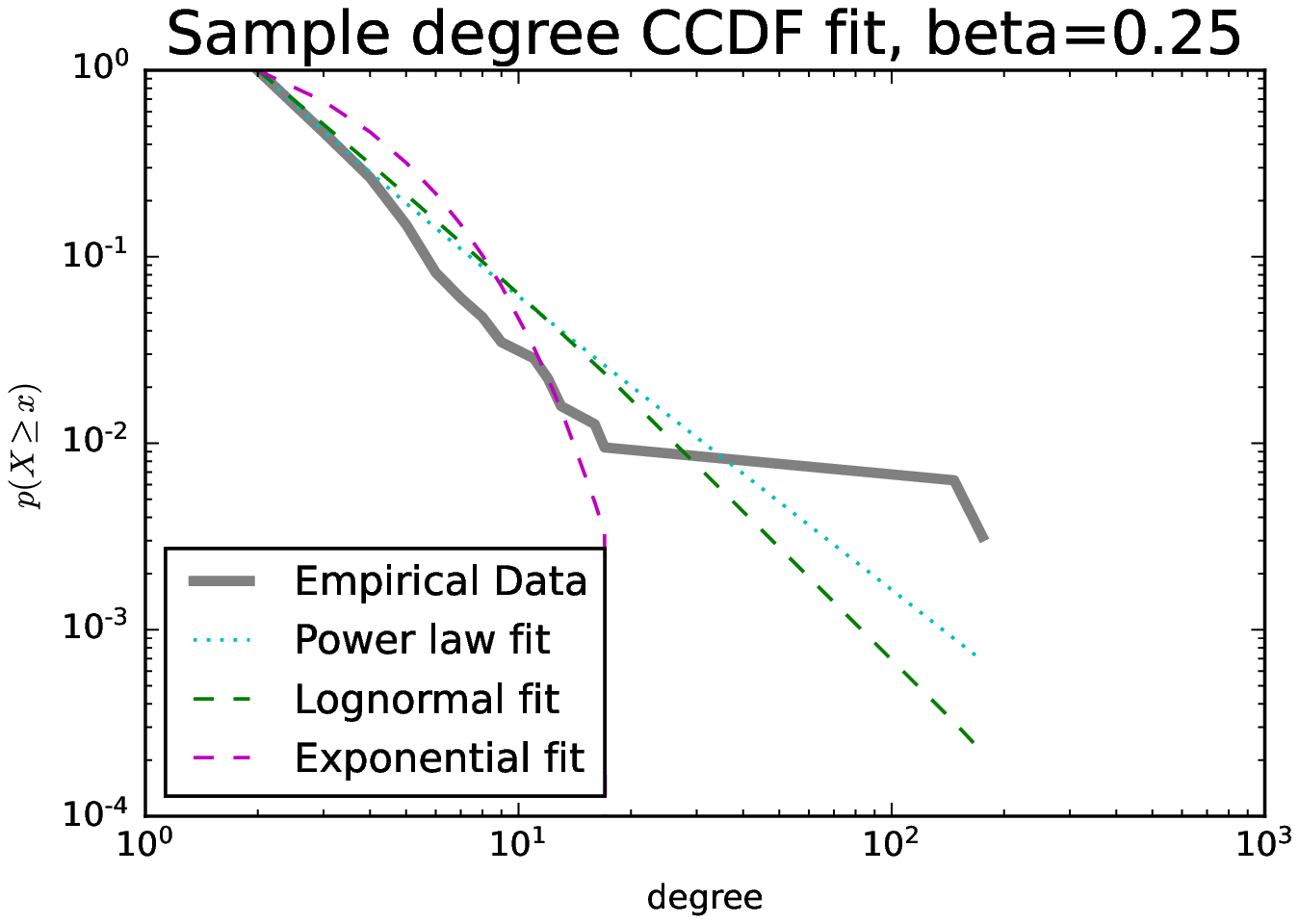}
%  \label{fig:rci_sub2}
%\end{subfigure}
%\setcounter{figure}{3}
\caption{This figure shows two sample degree CCDF fitting against power law and exponential distributions.}
\label{fig:fit}
\end{figure}

Table \ref{fit_table} show degree distribution fitness scores derived from the powerlaw package \cite{powerlaw}. For $\beta<0.3$, we find that a power law fit is better than an exponential distribution. However, the power law vs. log-normal distribution fit shows no strong tendency towards the power law. Fig. \ref{fig:fit} shows two sample distribution fits, for $\beta=0.8$ and $\beta=0.25$. These samples illustrate the tendency of the lower $\beta$ value to align with a log-normal distribution.

\begin{table}[tb]
\centering
 \caption{Experimental network degree distributions fitting against power law and exponential distributions, using log-likelihood ratios.}
 \begin{tabular}{c c c c c} 
 \hline
 $\beta$ & power vs. log-normal & p-value & power vs. exponential & p-value \\ [0.5ex] 
 \hline\hline
  0.1 &  -1.7128 & 0.5014 &  59.5303 & 0.0381 \\
  \hline
 0.2 &  -2.0651 & 0.4587 & 60.4613 & 0.0472 \\
  \hline
  0.3 &  -2.0798 & 0.3459 & 14.5674 & 0.2245 \\
  \hline
  0.4 &  -2.5701 & 0.2629 & -0.0826 & 0.5539 \\
  \hline
  0.5 &  -1.879 & 0.3331 & -1.2681 & 0.4705 \\
  \hline
   0.6 &  -1.6249 & 0.3299 & -1.3847 & 0.4336 \\
  \hline
  0.7 &  -1.8806 & 0.2915 & -1.5977 & 0.3776 \\
  \hline
  0.8 &  -2.2539 & 0.2546 & -1.8665 & 0.314 \\
  \hline
  0.9 &  -1.843 & 0.293 & -1.4889 & 0.3703 \\
  \hline
 1.0 &  -3.0878 & 0.2106 & -2.5583 & 0.2956 \\
 \hline
\end{tabular}
\label{fit_table}
\end{table}

\subsubsection{Influence of Recency Parameter.} 
\begin{figure}[tb]
\centering
 \includegraphics[width=.8\textwidth]{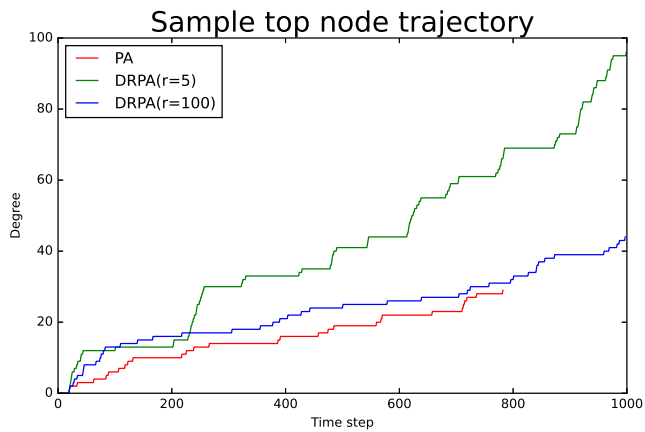}
 \caption{Sample runs ($\beta$=0.25 for DRPA) showing influence of $r$ values on node degree growth.}
 \label{fig:deg-traj}
\end{figure}

As we have seen, model parameter $\beta$ exerts an influence on both arrival order of high-ranking nodes as well as on the homogeneity of degree distribution. In contrast, parameter \emph{r} has little effect on the lateness of high-ranking late arrivals or degree distribution, but it does influence the degree trajectory of high-ranking nodes. 
Fig.\ \ref{fig:deg-traj} illustrates the influence of the parameter $r$ on the DRPA model. Lower $r$ tends to create bursts of degree growth, while larger $r$ smooths degree growth in a manner resembling PA. 

\section{Discussion}
There is a critical region ($\beta$ at 0.2-0.25) where we see both late-arriving-high-degree nodes and an overall network degree distribution that is close to a log-normal distribution (see table \ref{fit_table} and fig. \ref{fig:fit} in previous section). 

At lower $\beta$ values (but not close to 0) we observe a somewhat unusual network that has one or two extremely high-degree nodes and the rest are relatively homogeneous in terms of degree. One-off simulations with very low value of $\beta$ (e.g., 0.01) support this conjecture, showing that resulting networks are close to PA both in terms of late arrivals and overall degree distribution; however, those values were not included in the study's experimental runs. With $\beta$ value above the critical region, recency bias takes over leading to later high-degree nodes (see Fig.\ \ref{fig:rank_arrival}, left) as well as a relatively homogeneous degree distribution (see Table \ref{var_table} DRPA rows with higher $\beta$). This behavior makes sense because the stronger recency bias leads to hotness chasing, and little of the degree distribution shape from the PA model is left.

Experimental results are in partial agreement with analytical predictions proposed in Section 2, Eq.\ (\ref{eq:final}). The first two predictions are partially verified experimentally. As $\beta\to 0$, as previously discussed, one-off simulations show convergence with PA results. However, these parameter values were not included in the current study due to computational tractability, and follow up study is needed. Similarly, the prediction of long-term degree-based preference was not evaluated due to the size limitations on grown networks. %The prediction about $\beta\to 0$ is supported by experimental data. 
Finally, perhaps the most interesting analytical prediction is that there is a critical $\beta$ threshold that controls the impact of recency bias on network growth. While more study is needed to determine the exact value of this threshold, and to propose a theoretical causal mechanism, we tentatively propose that the region $\beta\approx 0.2 - 0.25$ is the general region of that threshold for the networks grown in the current study.

\section{Conclusion}

The DRPA model shows a variety of late-arrival patterns, both in terms of how late a node can arrive in a growing network and become a relatively high-degree node, as well as in terms of the grown network's degree distribution. We observed that the $\beta$ parameter is the primary parameter influencing these dynamics, and that parameter \emph{r} exerts a strong influence on the trajectory of degree change over a node's lifetime. 

Even though the DRPA model is a fairly simple extension of the PA model, the behavior we observe warrants future work. In particular, we suggest the following. First, the networks grown in this study were limited to 1,000 nodes, and to check analytical results regarding long-term behavior, much larger networks are needed. Second, more granular $\beta$ values will allow a deeper exploration of behavior regions that were identified in this study (e.g., emergence of super nodes at low-but-not-near-zero $\beta$ values). Finally, negative $\beta$ values were not used at all in the current study, but some one-off simulations suggest that they exert a preference for growth of slow-changing and high-ranking nodes, and thus another area of the DRPA model to study.


\begin{thebibliography}{4}

\bibitem{bianconi} Bianconi, G. and Barabási, A.L.: Competition and multiscaling in evolving networks. EPL (Europhysics Letters), 54(4), 436 (2001)

\bibitem{caldarelli} Caldarelli, G., Capocci, A., De Los Rios, P. and Munoz, M.A.: Scale-free networks from varying vertex intrinsic fitness. Physical review letters, 89(25), 258702 (2002)

\bibitem{fortunato} Fortunato, S., Flammini, A. and Menczer, F.: Scale-free network growth by ranking. Physical review letters, 96(21), 218701 (2006)

\bibitem{newman} Newman, M.E.: The first-mover advantage in scientific publication. EPL (Europhysics Letters), 86(6),  68001 (2009)

\bibitem{medo} Medo, M., Cimini, G. and Gualdi, S.: Temporal effects in the growth of networks. Physical review letters, 107(23), 238701 (2011)

\bibitem{liao} Liao, H., Mariani, M.S., Medo, M., Zhang, Y.C. and Zhou, M.Y.: Ranking in evolving complex networks. Physics Reports, 689, 1-54 (2017)

\bibitem{zhou} Zhou, Y., Zeng, A. and Wang, W.H.:. Temporal effects in trend prediction: identifying the most popular nodes in the future. PloS one, 10(3), 0120735 (2015)

\bibitem{candia} Candia, C., Jara-Figueroa, C., Rodriguez-Sickert, C., Barabási, A.L. and Hidalgo, C.A.: The universal decay of collective memory and attention. Nature human behaviour, 3(1),  82 (2019)

\bibitem{mokryn} Mokryn O, Wagner A, Blattner M, Ruppin E, Shavitt Y.: The Role of Temporal Trends in Growing Networks. PLoS ONE 11(8): e0156505  (2016)

\bibitem{lehmann} Lehmann, S., Jackson, A., Lautrup, B.: Life, Death and Preferential Attachment. Europhys. Lett., 69(2): pp. 298-303 (2005)

\bibitem{powerlaw} Alstott, J., Bullmore, E. and Plenz, D. powerlaw: a Python package for analysis of heavy-tailed distributions. PloS one, 9(1), p.e85777 (2014)

\end{thebibliography}
\end{document}